\documentclass[aip,apl,reprint]{revtex4-1}

\draft 
\usepackage{amssymb}
\usepackage{graphicx}
\usepackage{dcolumn}
\usepackage{bm}
\usepackage{color}
\usepackage{float}
\usepackage{hyperref}
\usepackage[version=4]{mhchem}
\usepackage{relsize}

\begin{document}

\title{Beamed UV sonoluminescence by aspherical air bubble collapse near liquid-metal microparticles}

\author{Bradley Boyd}
\affiliation{Department of Mechanical Engineering, University of Canterbury, 20 Kirkwood Ave, Upper Riccarton, Christchurch 8041, New Zealand}

\author{Sergey A.~Suslov}
\affiliation{Department of Mathematics, Swinburne University of Technology, Hawthorn, Victoria 3122, Australia\looseness=-1}

\author{Sid Becker}
\affiliation{Department of Mechanical Engineering, University of Canterbury, 20 Kirkwood Ave, Upper Riccarton, Christchurch 8041, New Zealand}

\author{Andrew D.~Greentree}
\affiliation{Australian Research Council Centre of Excellence for Nanoscale BioPhotonics, School of Science, RMIT University, Melbourne, Victoria 3001, Australia}
 
\author{Ivan S.~Maksymov}
\email{imaksymov@swin.edu.au}
\affiliation{Centre for Micro-Photonics, Swinburne University of Technology, Hawthorn, Victoria 3122, Australia\looseness=-1}


\begin{abstract}

Irradiation with UV-C band ultraviolet light is one of the most commonly used ways of disinfecting water contaminated by pathogens such as bacteria and viruses. Sonoluminescence, the emission of light from acoustically-induced collapse of air bubbles in water, is an efficient means of generating UV-C light. However, because a spherical bubble collapsing in the bulk of water creates isotropic radiation, the generated UV-C light fluence is insufficient for disinfection. Here, we show that we can create a UV light beam from aspherical air bubble collapse near a gallium-based liquid-metal microparticle. The beam is perpendicular to the metal surface and is caused by the interaction of sonoluminescence light with UV plasmon modes of the metal. We calculate that such beams can generate fluences exceeding $10$\,mJ/cm$^2$, which is sufficient to irreversibly inactivate most common pathogens in water with the turbidity of more than $5$\,Nephelometric Turbidity Units.

\end{abstract}

\maketitle 


\textit{Introduction.}---The ability of UV-C light ($200-280$\,nm) to inactivate bacteria, viruses and protozoa is widely used as an environmentally-friendly, chemical-free and highly effective means of disinfecting and safeguarding water against pathogens responsible for cholera, polio, typhoid, hepatitis and other bacterial, viral and parasitic diseases \cite{Bol08}. UV-C light inactivates pathogens through absorption of radiation energy by their cellular RNA and DNA prompting the formation of new bonds between adjacent nucleotides. This results in a photochemical damage that renders pathogens incapable of reproducing and infecting \cite{Bol08}.

However, some pathogens can recover from photochemical damage when the initial UV dosage (fluence) is not sufficiently high \cite{Bol08}. Thus, the fluence must exceed $5$ and $10$\,mJ/cm$^2$, respectively, to inactivate $99\%$ and $99.9\%$ of \textit{Giardia} and \textit{Cryptosporidium} pathogens \cite{WHO}. These specifications are for water purified from solid particles larger than $5-10$\,$\mu$m [turbidity less than $5$\,Nephelometric Turbidity Units (NTU)] \cite{WHO}. Otherwise, particles can shield pathogens from the UV light, thereby allowing many pathogens to recover and infect. 
\begin{figure}[t]
\centerline{
\includegraphics[width=8.5cm]{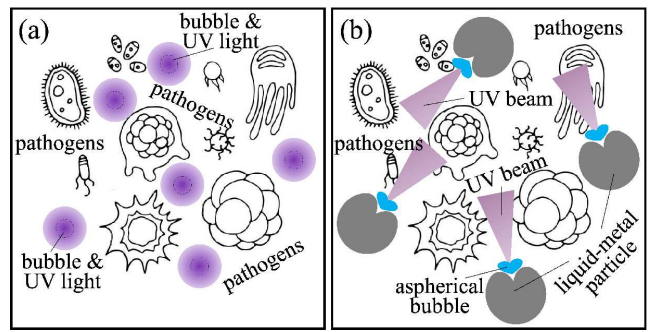}}
\caption{(a)~The collapse of a single spherical air bubble in water creates isotropic, low-fluence radiation unsuitable for UV germicidal irradiation, (b)~the collapse of an aspherical bubble near a liquid-metal particle results in the deformation of the liquid-metal surface and interaction of the emitted light with UV plasmon modes of the microparticle. This creates directed, high-fluence UV-C light beams capable of inactivating $99.9\%$ of most common pathogens.}
\label{fig:fig1}
\end{figure}

The filtration of natural water presents significant challenges for remote communities and developing nations \cite{WHO}. Moreover, filtered water, dissolved iron, organic salts and the pathogen population itself absorb UV-C light. Therefore, a $50\%$ UV radiation loss has been accepted as suitable for practical use \cite{Bol08}.  

To enable UV disinfection of turbid water, we use the effect of sonoluminescence---the emission of broadband UV light in acoustically-induced collapse of air bubbles in water \cite{Put00, Bre02}. Air bubbles suitable for sonoluminescence can be created in natural water \cite{Tem17} and we show that they can act as compact sources of germicidal radiation located several optical wavelengths away from pathogens. This means that shielding of pathogens by particles suspended in water would be greatly reduced, but small distances travelled by UV-C light between the source and pathogens would result in negligible absorption losses. 

However, conventional sonoluminescence of spherical air bubbles produces isotropic radiation \cite{Put00, Bre02}, which means that the fluence of light generated by any single collapse event is low compared to that required for UV germicidal irradiation [Fig.~\ref{fig:fig1}(a)]. We demonstrate that it is possible to create a directed UV-C light beam via sonoluminescence of air bubbles collapsing near microparticles made of non-toxic and environmentally friendly gallium-based alloys \cite{Dic08}. The resulting beam is perpendicular to the metal surface because of the asphericity of the bubble collapse and the resonant interaction of the emitted light with UV-C plasmons in the microparticle \cite{Rei18} [Fig.~\ref{fig:fig1}(b)]. We demonstrate that such beams are capable of generating UV-C fluences exceeding the thresholds required to irreversibly inactivate $99.9\%$ of most common pathogens. Because the melting point of gallium-alloy metals is smaller than room temperature \cite{Dic08, Sur05}, we also show that the germicidal effect can be achieved with both liquid and solid metal microparticles.

\textit{Spherical bubble collapse in unbounded water.}---The collapse of a spherical bubble in unbounded liquid is described by the bubble radius $R(t)$ that oscillates around the equilibrium radius $R_0$. We employ a Rayleigh-Plesset model that is commonly used in the context of sonoluminescence \cite{Bre02}: $\rho(R\ddot{R} {+} \frac{3}{2}\dot{R}^2){=}[p_g{-}p_0{-}P(t)]{-}4\eta\dot{R}/R{-}2\sigma/R{+}Rv^{-1}\frac{dp_g}{dt}$, where $\dot{R}$ and $\ddot{R}$ are the first and second time derivatives of $R(t)$, $p_g(t){=}(p_0{+}\frac{2\sigma}{R})\frac{(R_0^3-a^3)^\gamma}{[R(t)^3-a^3]^\gamma}$, $P(t)=-p_{\max}\sin(2\pi ft)$ with the frequency $f$ and time $t$, and $p_0=101.3$\,kPa. We solve this equation by using a fourth order Runge-Kutta scheme \cite{Mak17}. The material parameters are \cite{Mak17, Bre02}: density $\rho=1000$\,kg/m$^3$, sound velocity $v=1500$\,m/s, viscosity $\eta=1.002\times10^{-3}$\,Pa\,s, surface tension $\sigma=0.0728$\,N/m, $\gamma=1.4$ and $a=R_0/8.5$. The temperature inside the bubble is $T(t){=}T_0\frac{(R_0^3-a^3)^{(\gamma-1)}}{[R(t)^3-a^3]^{(\gamma-1)}}$ with $T_0{=}298$\,K (Ref.~\onlinecite{Bre02}).

Sonoluminescence can be observed in a limited range of bubble parameters that enable gas exchange processes, high temperatures and bubble shape stability \cite{Bre02}. Atmospheric air contains ${\sim}1\%$ of argon and high-temperature sonochemical reactions inside the collapsing bubble lead to a rapid removal of oxygen and nitrogen thereby leaving only the noble gas inside the bubble and enabling the emission of light \cite{Bre02}. According to the Blake threshold criterion for sonoluminescence \cite{Bre02}, $R_0$ has to be greater than $R^{C}_0 =C\sigma/(p_{\max}-p_0)$. With $C\approx0.77$ and $p_{\max}=1.7p_0$ (Ref.~\onlinecite{Bre02}), we obtain $R^{C}_0\approx0.8$\,$\mu$m. Hence, in the following we assume that $R_0=1\,\mu$m. We also ensure that $f<f_0$, where $f_0$ is the main resonance frequency of the air bubble in water ($f_0R_0\approx3.26$\,m/s).
\begin{figure}[t]
\centerline{
\includegraphics[width=8.0cm]{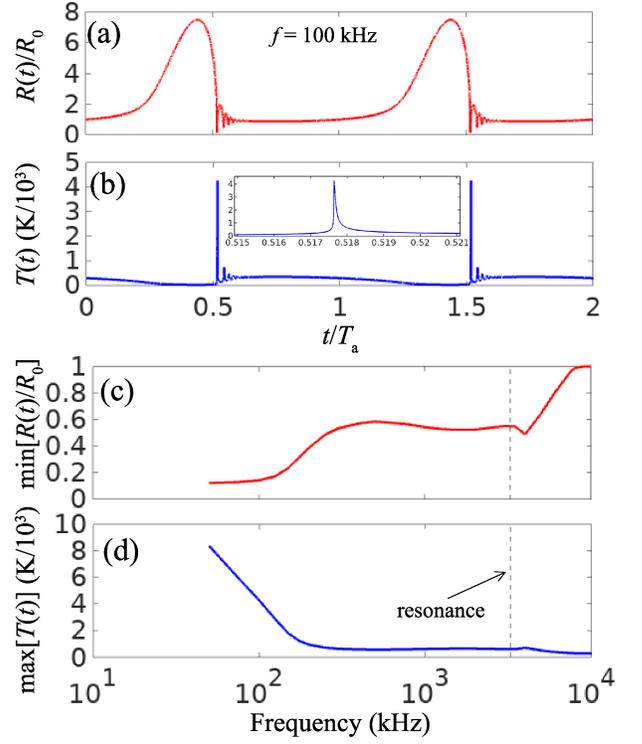}}
\caption{(a)~Calculated non-dimensional bubble radius $R(t)/R_0$ ($R_0{=}1\,\mu$m) and (b)~temperature $T(t)$ of the bubble as a function of time $t$ given in units of the acoustic wave period $T_a$. The acoustic wave frequency is $f{=}100$\,kHz and the peak pressure amplitude is $p_{\max}{=}1.7p_0$. (c)~Minimum bubble radius $\min[R(t)/R_0]$ and (d)~peak temperature $\max[T(t)]$ of the bubble as functions of $f$. The broken vertical lines show the main bubble resonance frequency $f_0$. }
\label{fig:fig2}
\end{figure}

As a representative example, Fig.~\ref{fig:fig2}(a) shows the dynamics of the non-dimensional bubble radius $R(t)/R_0$ for $f=100$\,kHz. We observe a large excursion of $R(t)$ from $R_0$ followed by a steep collapse with a series of sharp afterbounces. The temperature of the bubble dramatically increases when $R(t)/R_0$ reaches its minimum [Fig.~\ref{fig:fig2}(b)], which results in the sonoluminescence pulse. The duration of this pulse is defined by the full width at half maximum (FWHM) of the temperature peak \cite{Put00}. The minimum radius $\min[R(t)/R_0]$, the peak temperature $\max[T(t)]$ of the bubble and the FWHM of the light pulse depend on the acoustic frequency $f$ [Fig.~\ref{fig:fig2}(c, d)]. The value of $\max[T(t)]$ varies slowly when $f>f_0$, remains approximately constant for $f\approx f_0$ and increases quickly at $f<f_0$ when the collapse becomes more violent.    
\begin{figure*}[t]
\centerline{
\includegraphics[width=18cm]{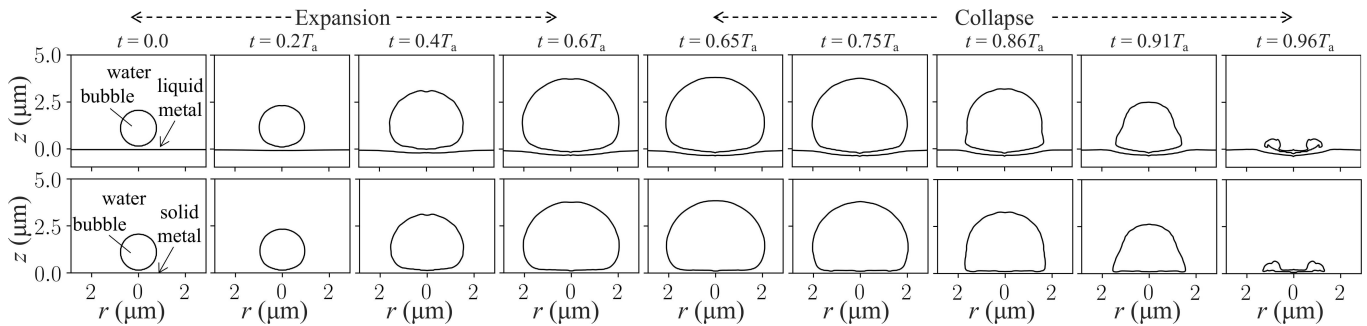}}
\caption{Representative axisymmetric profiles of an air bubble during its expansion and collapse near the liquid metal (top row) and solid metal (bottom row). $T_{\rm{a}}$ is the period of the sinusoidal acoustic wave [$f{=}1.5$\,MHz, $p_{\max}{=}2p_0$ (Ref.~\onlinecite{Boy18})] incident along the $z$-axis toward the metal surface. A concave shape assumed by the liquid-metal surface as a result of the collapse will allow us to focus sonoluminescence light into a more intense beam compared to that near a flat solid-metal surface.}
\label{fig:fig3}
\end{figure*}

\textit{Aspherical bubble collapse near metal microparticles.}---When the bubble oscillates near a solid surface, it flattens near the surface \cite{Fon06, Cur13, Boy18, Boy19}. At the collapse stage, a water jet develops through the centre of the bubble toward the surface. The impact of the jet on the surface results in large pressure that can potentially cause damage \cite{Fon06, Cur13, Boy18, Boy19}. The collapse near a solid surface also results in stable sonoluminescence \cite{Wen97, Put00}.

We analyse an aspherical bubble collapse near spherical gallium-alloy particles of $50-100\,\mu$m radius, which can be fabricated by means of a self-breakup of a liquid metal jet \cite{Dic08}. We assume that an initially spherical bubble with $R_0 = 1\,\mu$m nearly touches the particle---the distance between the bubble centre and the metal surface is $1.1\,\mu$m. Because $R_0$ is $50-100$ times smaller than the radius of the particle, we simplify our model by considering a planar $100\,\mu$m-thick liquid-metal layer.

We use a high-order numerical method for solving fully-compressible multiphase inviscid flow equations \cite{Boy18, Boy19, thesis}: $\partial_t(\alpha_i\rho_i)+\nabla\cdot(\alpha_i \rho_i\bm{u})=0$ ($i=1,2,3$), $\partial_t(\rho \bm{u})+\nabla\cdot(\rho\bm{u}\otimes\bm{u}+p\bm{I})=0$, $\partial_tE+\nabla \cdot[\bm{u}(E+p)]=0$, $\partial_t\alpha_1+\bm{u}\cdot\nabla\alpha_1=0$, and $\partial_t(\alpha_1+\alpha_2)+\bm{u}\cdot\nabla(\alpha_1+\alpha_2)=0$, where $\alpha_i$ are the volume fractions of bubble gas, water, and liquid metal ($\sum_i\alpha_i=1$), $\rho=\sum_i \alpha_i \rho_i$ is the density, $\bm{u}$ is the velocity vector, $p$ is the pressure, $E=\sum_i \alpha_i E_i$ is the total energy, and $\bm{I}$ is the identity matrix. 

These equations combined with an equation of state (EOS) for the mixture of bubble gas, water, and liquid metal define the compressible-multiphase system \cite{thesis}. We neglect viscous forces because they are significantly smaller than the pressure forces driving the collapse. The influence of surface tension is also neglected because the primary force driving the collapse is due to the difference between the pressure inside the bubble and the acoustic pressure \cite{thesis}. The density of the liquid metal is 6360\,kg/m$^3$ and the parameters for the EOS are taken from Ref.~\onlinecite{ref2}. The presence of the $1-3$\,nm-thick gallium oxide layer \cite{Rei18} is neglected because it does not cause qualitative changes in fluid-mechanical and optical properties of the metal.

Figure~\ref{fig:fig3} shows the profiles of the bubble and metal surface during one period $T_{\rm{a}}$ of the $f=1.5$\,MHz sinusoidal acoustic wave that triggers the expansion-collapse cycles of the bubble and the deformation of the liquid-metal surface. A toroidal bubble is formed in the end of the collapse stage, which is also the instance when the sonoluminescence light is emitted \cite{Wen97, Put00}. This behaviour closely resembles that of the bubble located near the solid metal surface, but the solid metal is not deformed.

The formation of a toroidal bubble is accompanied by a water micro-jet impinging the metal surface \cite{Fon06, Cur13, Boy18, Boy19}. We calculate the pressure developed by the micro-jet near the \textit{solid metal} to be $\sim7$\,MPa, which agress with experimental data \cite{Fon06, Cur13}. This micro-jet has the potential to damage the solid metal surface \cite{Cur13}. Our model does not allow us to account for this effect, but it is well-known that the so-created roughness of the metal surface can lead to undesirable scattering and absorption of light.

When modelling the motion of the \textit{liquid-metal} surface, we assume that due to a periodicity of the bubble expansion-collapse process it would regain its initial shape in the beginning of each cycle. We predict the jet pressure at the liquid metal surface to be $\sim26$\,MPa. This value may be somewhat overestimated because of a numerical instability arising in the end of the collapse stage, but experimental data suitable for validation of this prediction are currently not available.

The numerical instabilities grow as the frequency $f$ decreases because the collapse becomes more violent. Thus, we follow the theoretical prediction \cite{Wan10} of consistent aspherical collapse shapes at $f{<}f_0{\approx}3.26$\,MHz. Similar to the radius of spherical bubbles $R(t)$ (Fig.~\ref{fig:fig2}), the volume of aspherical bubbles $V(t)$ exhibits larger excursions from $V_0$ followed by steeper collapses when $f$ is decreased \cite{Wan10}. This enables us to use the frequency dependence of the temperature of spherical bubbles [Fig.~\ref{fig:fig2}(d)] to predict the temperature of aspherical ones.

\textit{Model of sonoluminescence.}---To simulate the emission of the sonoluminescence light, we use a three-dimensional finite-difference time-domain (FDTD) method based on Maxwell's equations \cite{Mak17}. The fluid-mechanical and FDTD models have two very different time scales associated with a low acoustic frequency $f$ and high UV light frequency. This means that the optical solution reaches a quasi-steady state before the shapes of the bubble and liquid metal have changed substantially. Thus, in the FDTD model we can use the instantaneous snapshots of the bubble and liquid-metal profiles from Figs.~\ref{fig:fig2} and~\ref{fig:fig3}.

We use the blackbody emission model \cite{Bre02} to simulate the creation of sonoluminescence light by a collapsing bubble \cite{Bre02}. A blackbody of temperature $T$ produces a spectral radiance $L_\lambda[T]=2h c^2\lambda^{-5} [\exp(h c/\lambda k_{\rm{B}} T)-1]^{-1}$ with the Planck and Boltzmann constants $h\approx6.63\times10^{-34}$\,m$^2$kg/s and $k_{\rm{B}}\approx1.38\times10^{-23}$\,m$^2$kg/(s$^2$K), and the speed of light in  vacuum $c\approx3\times10^8$\,m/s. The spectral radiant power is calculated by integrating $L_\lambda$ over the projected bubble surface and all solid angles \cite{Bre02}. For a spherical bubble it is $\Phi_\lambda(t)=4\pi^2R(t)^2L_\lambda[T(t)]$ with $R(t)$ and $T(t)$ drawn from Fig.~\ref{fig:fig2}.

The FDTD method can readily simulate wide-spectrum signals such as $L_\lambda$. However, it does not allow for a quantitative control of the emitted power, which means that output energy quantities are expressed in arbitrary units. To calculate the fluence in real physical units, we exploit the linearity of Maxwell's equations and first simulate the spatial pattern of the emitted light. This pattern corresponds to the profile of radiant emittance, a radiometric term that is equivalent to intensity in optics, expressed in arbitrary units. Then, we semi-analytically calculate the total amount of the radiant power as $\Phi(t)=\int_{\lambda_1}^{\lambda_2} \Phi_\lambda(t)\,{\rm d}\lambda$ with $\lambda_1=200$\,nm and $\lambda_2=280$\,nm. The time integration of $\Phi(t)$ gives the optical energy in Joules, but the spatial pattern of the emitted light allows us to define the area through which the radiation passes. This allows us to calculate the fluence as the energy delivered per unit area.     
\begin{figure}[t]
\centerline{
\includegraphics[width=8.5cm]{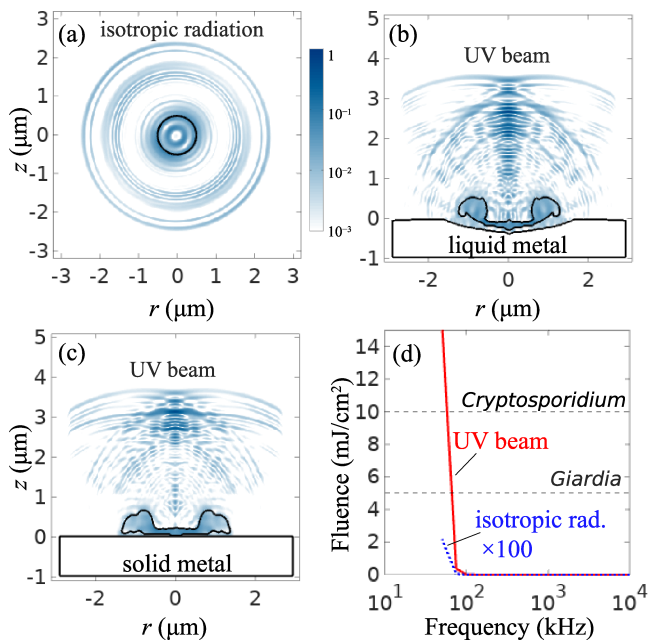}}
\caption{Spatial patterns of light emitted by (a)~spherical bubble in unbounded water, (b)~aspherical bubble near the \textit{liquid-metal} surface, and (c)~aspherical bubble near the \textit{solid-metal} surface. All patterns are normalised to the maximum magnitude in (b). (d)~UV radiation fluence as a function of the acoustic frequency $f$. The fluence of isotropic radiation is multiplied by $100$. The fluence required for the inactivation of \textit{Giardia} and \textit{Cryptosporidium} pathogens is indicated.}
\label{fig:fig4}
\end{figure}

The spatial resolution of the FDTD mesh is $5$\,nm. The optical refractive index of water in the UV-C band is $1.366$ (Ref.~\onlinecite{Hal73}). The complex dielectric permittivity of the gallium alloy is described by a Drude-like model \cite{Rei18}.

Figure~\ref{fig:fig4} shows the steady-state spatial pattern of sonoluminescence light at $f{=}1.5$\,MHz. The radius of the spherical bubble is $500$\,nm [Fig.~\ref{fig:fig2}(c)] and the aspherical profiles are taken from Fig.~\ref{fig:fig3}. By virtue of the Fourier transform, this time-domain wave packet carries the spectral radiance $L_\lambda$ for $\lambda=200-280$\,nm. A spherical bubble produces isotropic radiation while its aspherical counterpart produces a beam near both liquid and solid metal surfaces. The beam arising near a deformed liquid-metal surface acting as a concave mirror is more intense compared to that near a flat solid surface.

The energetics of sonoluminescence produced by spherical and aspherical bubbles are very similar \cite{Wen97, Put00}, which in the framework of the blackbody model requires $T(t)$ to follow the same trend. Hence, we use $T(t)$ from Fig.~\ref{fig:fig2} to calculate the UV light fluence generated by both spherical and aspherical bubbles. For the spherical bubble, we use the area of an imaginary sphere with the radius $R_0=1\,\mu$m. For the aspherical bubble, we take the radius ($\sim75$\,nm) and the cross-sectional area of the beam from the spatial profile in Fig.~\ref{fig:fig3}(b).

Figure~\ref{fig:fig4}(d) shows that the fluence generated by both spherical and aspherical bubbles dramatically increases at $f<100$\,kHz. However, only the formation of a directed beam near the metal microparticles results in the irreversible inactivation of pathogens. The improvement of the germicidal effectiveness is due to the sharp increase in temperature $T$ as $f$ decreases [Fig.~\ref{fig:fig2}(d)]. The wavelength of the peak of the blackbody radiation scales as $\lambda_p=b/T$ with $b\approx2.9\times10^{-3}$\,K\,m, which implies that $\lambda_p$ shifts to the UV-C band when $f$ is decreased.   

\textit{Conclusions.}---We have demonstrated that the aspherical collapse of gas bubbles and sonoluminescence near liquid-metal particles result in the generation of UV light beams capable of inactivating pathogens contaminating drinking water. A water disinfection system based on this scheme should be affordable for remote communities and developing nations. It would be suitable for disinfecting turbid water and could be combined with water treatment systems using bubbles \cite{Tem17}. 

This work was supported by the Australian Research Council (ARC) through the Future Fellowship (FT180100343, FT160100357) and LIEF (LE160100051) programs, by the Centre of Excellence for Nanoscale BioPhotonics (CE140100003) and by the Royal Society of New Zealand’s Marsden Fund (UOC1305).

%

\end{document}